\documentclass[prb,superscriptaddress,twocolumn,showpacs,preprintnumbers,amsmath,amssymb]{revtex4}

\usepackage{graphicx}
\usepackage{dcolumn}
\usepackage{bm}

\newcommand{\uh}{\textrm{h}}

\DeclareBoldMathCommand{\bnabla}{\nabla}
\DeclareBoldMathCommand{\bcdot} {\cdot}
\DeclareBoldMathCommand{\btimes}{\times}

\newcommand{\ud}[1]{\textrm{d}#1\,}

\begin{document}

\preprint{APS/123-QED}

\title{Theoretical derivation of the bi- and tri-molecular trion formation coefficients}
\author{J.~Berney}
 \affiliation{Institut de Photonique et Electronique Quantiques,
 Ecole Polytechnique F\'{e}d\'{e}rale de Lausanne (EPFL) CH1015
 Lausanne, Switzerland}
  \affiliation{Attolight S\`{a}rl, Chemin de la Raye 13, CH1024 Ecublens, Switzerland}
\author{M.T.~Portella-Oberli}
 \affiliation{Institut de Photonique et Electronique Quantiques,
 Ecole Polytechnique F\'{e}d\'{e}rale de Lausanne (EPFL) CH1015
 Lausanne, Switzerland}
\author{B.~Deveaud-Pl\'{e}dran}
 \affiliation{Institut de Photonique et Electronique Quantiques,
 Ecole Polytechnique F\'{e}d\'{e}rale de Lausanne (EPFL) CH1015
 Lausanne, Switzerland}
\date{\today}

\begin{abstract}
A theoretical investigation of the trion formation process from free carriers in a single GaAs/ Al$_{1-x}$Ga$_x$As quantum well is presented. The mechanism for the formation process is provided by the interaction of the electrons and holes with phonons. The contributions from both the acoustic and optical phonons are considered. The dependence of both bi-molecular and tri-molecular formation rates on temperature is calculated. We demonstrate that they are equivalent for negatively and positively charged excitons.\end{abstract}


\maketitle

\section{Introduction}

In semiconductor wells, the photoluminescence spectrum following the generation of electron hole plasma is dominated by an exciton line \cite{weisbuch:1981, deveaud:1987, yoon:1996, damen:1990,robart:1995,deveaud:1993}. The formation of excitons in quantum wells has been extensively investigated both experimentally  \cite{damen:1990,robart:1995,deveaud:1993,kaindl:2003,szczytko:2004,szczytko:2005} and theoretically \cite{thilagam:1993, piermarocchi:1996,piermarocchi:1997}. Recently, it was experimentally showed that the exciton formation is strongly density and temperature dependent; it is a bimolecular process in which an electron and a hole are bound by Coulomb interaction with the emission of the appropriate phonon \cite{szczytko:2004}. This experimental result unambiguously confirmed the theoretical prediction for luminescence spectrum is modified: a charged exciton resonance appears below that of the exciton. 

In a previous publication \cite{portella:2009}, we showed experimentally that the dynamics of exciton, trion and electron-hole plasma can be ruled by a simple rate equation model, in which we account for
bimolecular formation of excitons from an electron-hole plasma, bimolecular
formation of trions from excitons and free carriers and trimolecular formation
from free carriers. Using only two fit parameters, we were able to extract experimentally the
dependence of both bi-molecular and tri-molecular formation
coefficients on temperature.

In this paper, we propose a theoretical derivation of these coefficients. We consider formation channels through which  the formation channel  and show that they correspond to experimental calculations.

In Sec. I, we first reproduce the results of
\citet{piermarocchi:1997} on exciton formation. We then extend the
formalism to the case of charged excitons; in Sec. II, we calculate the bi-molecular formation and in Sec. III the tri-molecular formation. In Sec. IV, we show how these formation rates can be used to calculate formation dynamics at thermodynamical equilibrium.

\section{Bimolecular formation of excitons}

The dynamics of the exciton formation is considered in the framework
of the Boltzmann equation for a system containing free electrons,
free holes, and excitons. The residual Coulomb interaction between
the free carriers is neglected, which is justified in the range of
temperatures and densities considered \cite{portella:2009}. In this work we focus on the
exciton formation mechanism and do not discuss the relaxation of the
three species within their respective bands, the electron-hole
scattering, and radiative recombination. We denote the occupation
numbers for electrons, holes, and excitons by $f_e(\bm{k}_e)$,
$f_h(\bm{k}_h)$, and $f_X(\bm{k}_X)$, respectively, where
$\bm{k}_e$, $\bm{k}_h$, and $\bm{k}_X$ are the in-plane momenta for
electrons, holes, and excitons. For the bimolecular formation,
following \citet{piermarocchi:1997}, the scattering terms in the
Boltzmann equation process reads
\begin{equation} \label{eq:bimol_form}
\left( \frac{\ud f_e(\bm{k_e})}{\ud t}
\right)_{form}=-\sum_{\bm{k_X},\bm{k_h}}
w_{\bm{k_e},\bm{k_h}\to\bm{k_{X}}} f_h(\bm{k_h}) f_e(\bm{k_e}),
\end{equation}
where $w_{\bm{k_e},\bm{k_h}\to\bm{k_{X}}}$ represents the
probability per unit time for a free electron and a free hole to
bind together and form an exciton. Free carriers thermalize very
quickly in comparison to the exciton formation time, notably through
fast carrier-carrier scattering \citep{knox:1992}. It is thus
assumed that during the evolution of the system, the free electrons
and holes are thermalized at the same temperature $T_c$. In the
scattering term of Eq.~\eqref{eq:bimol_form}, we use for $f_e(k_e)$,
and $f_h(k_h)$ equilibrium Boltzmann distribution function at $T_c$.
Consequently, by summing Eq. \eqref{eq:bimol_form} over $\bm{k_e}$,
we obtain an adiabatic equation for the evolution of the electronic
density $n_e=\frac{1}{S} \sum_{\bm{k_e}} f_e(\bm{k_e})$
\begin{equation} \label{eq:bimol_form_adiab}
\frac{\ud n_e}{\ud t}=-\sum_{\bm{k_{X}}} F(\bm{k_{X}}) n_e n_h
\equiv -C n_e n_h.
\end{equation}
The coefficient $C$ is the bimolecular formation coefficient, which
depends on both $T_c$ and the lattice temperature $T_l$ through the
term
\begin{multline} \label{eq:coef_C}
F(\bm{k_{X}})=\left( \frac{2 \pi \hbar^2}{k_BT_c} \right)^2
\frac{1}{m_e m_h S} \\
\times \sum_{\bm{k_e},\bm{k_h}}
w_{\bm{k_e},\bm{k_h}\to\bm{k_{X}}
}e^{-(E_e(\bm{k_e})+E_h(\bm{k_h}))/k_BT_c},
\end{multline}
where $S$ denotes the QW surface area.

Free carriers are coupled to the exciton by a continuum of phonon
states $(\bm{q}, q_z)$ through a carrier-phonon interaction
Hamiltonian $\mathcal{H}_{e/h-ph}$. A phonon can be emitted~(+) or
absorbed~(-) in the formation process of the exciton. We calculate
both case separatly using the Fermi's golden rule\\
\begin{multline}\label{eq:fermi_golden_rule}
w^{\pm}_{\bm{k}_e,\bm{k}_h \to \bm{k}_{X}} = \frac{2\pi}{\hbar}
\sum_{\bm{q},q_z} \\ \left| {\langle\bm{k}_X | \otimes \langle
n_{\bm{q},q_z}\pm1|\, \mathcal{H}_{e/h-ph} \, |n_{\bm{q},q_z}
\rangle \otimes |\bm{k}_e \rangle \otimes |\bm{k}_h \rangle}
 \right|^2\\
 \times \delta\big[ E_e(\bm{k}_e)+E_h(\bm{k}_h)-E_X(\bm{k}_X)\mp\hbar
 \omega_{ph}(\bm{q},q_z)\big],
\end{multline}
with $E_e(\bm{k}_e)$, $E_h(\bm{k}_h)$ and $E_X(\bm{k}_X)$ the energy
dispertion of the electrons, holes and excitons respectively, and
$\hbar \omega_{ph}(\bm{q},q_z)$ the energy of the emitted (absorbed)
phonon. We first build the bound and unbound electron-hole pair
states $|\bm{k}_X \rangle$ and $|\bm{k}_e \rangle \otimes |\bm{k}_h
\rangle$.

\subsection{Bound and unbound exciton states}
Let $(\bm{r}_{e\parallel},z_e)$ and $(\bm{r}_{h\parallel},z_h)$
be the electron and hole position vectors respectively and
$\Phi^X[(\bm{r}_{e\parallel},z_e),(\bm{r}_{h\parallel},z_h)]$ the
exciton wavefunction, where we have separated the coordinates in the
QW plane (x-y) from the perpendicular coordinates~(z). Denoting the
electron (hole) in-plane momenta $\bm{k}_e$ ($\bm{k}_h$), we write
the in-plane Fourier transform of this function
\begin{align}\label{eq:X_wavefct_abs_coor}
\Phi^X_{\bm{k}_e,\bm{k}_h}(z_e,z_h)=&\frac{1}{S}\int
\ud\bm{r}_{e\parallel}\ud \bm{r}_{h\parallel}\Phi^X[(\bm{r}_{e\parallel},z_e),(\bm{r}_{h\parallel},z_h)]\nonumber\\ 
&\times e^{-i(\bm{k}_e\cdot\bm{r}_{e\parallel}+\bm{k}_h\cdot\bm{r}_{h\parallel})},
\end{align}
where $S$ denotes the QW surface area. Transforming to
center-of-mass (CM) and relative coordinates in the QW plane ---
$\bm{R}_\parallel=\alpha_X
\bm{r}_{e\parallel}+\beta_X\bm{r}_{h\parallel}$,
$\bm{r}_\parallel=\bm{r}_{e\parallel}-\bm{r}_{h\parallel}$, where
$\alpha_X=m_e/M_X$, $\beta_X=m_h/M_X$ and $m_e$, $m_h$, $M_X$ are
the electron, hole and exciton in-plane effective mass --- we can
apply Bloch's theorem and decompose the exciton wavefunction into a
free motion part $e^{i\bm{k}_X \cdot \bm{R}_\parallel}$ related to
the exciton in-plane momentum $\bm{k}_X$ and an envelope function.
To facilitate the calculation, we use an envelope function separable
in  $z$ and $r_\parallel$, although it is strictly justifiable only
for narrow well structures,
\begin{equation}
 \phi^X(r_\parallel,z_e,z_h)=\chi_e(z_e)\chi_h(z_h)\varphi^{\lambda_X}(r_\parallel).
\end{equation}
The confinement functions $\chi_e(z_e)$ ($\chi_h(z_h)$) is taken to
be the wavefunction of an electron (hole) in the ground state of a
finite square quantum well \citep{bastard:1988}
\begin{equation}\label{eq:finitesquareqw}
\chi_\alpha(z_\alpha) = \left\{
\begin{array}{ll}
A_\alpha\cos\left(k_{z}^{w(\alpha)} z_\alpha\right) & \textrm{for}\quad |z_\alpha|<\frac{L_z}{2}\\
B_\alpha\exp\left(-k_{z}^{b(\alpha)}(|z_\alpha|-L_z/2)\right) &
\textrm{for}\quad |z_\alpha|>\frac{L_z}{2}
\end{array}\right.
\end{equation}
with $\alpha=e,h$ and we use the simplest electron orbital function
\begin{equation}\label{eq:1s_wavefct}
\varphi^{\lambda_X}(r_\parallel)=\sqrt{\frac{2}{\pi
\lambda_X^2}}e^{-r_\parallel/\lambda_X},
\end{equation}
whose in-plane Fourier transform is given by
\begin{equation}\label{eq:1s_fourier}
    \varphi^{\lambda_X}_{\bm{k}}=\sqrt{\frac{8\pi{\lambda_X}^2}{S}}\left[1+({\lambda_X} k)^2\right]^{-3/2}.
\end{equation}
The variational parameter $\lambda_X$ is associated with the Bohr
radius of the exciton in the QW. Eq.~\eqref{eq:X_wavefct_abs_coor}
can be rewritten as
\begin{align}\label{eq:X_wavefct_rel_coor}
\Phi^X_{\bm{k}_e,\bm{k}_h}(z_e,z_h)=&\frac{1}{S}\int \ud\bm{R}_{\parallel}\ud \bm{r}_{\parallel} \phi^X(r_\parallel,z_e,z_h)\notag\\
&\times e^{i\bm{k}_X\cdot\bm{R}_\parallel} e^{-i[\bm{R}_\parallel\cdot(\bm{k}_e+\bm{k}_h)+\bm{r}_{\parallel}\cdot(\beta_X\bm{k}_e-\alpha_X\bm{k}_{h})]}\notag\\
=&\,\delta_{\bm{k}_X-\bm{k}_e-\bm{k}_h}
\phi^X_{\bm{k}_e-\alpha_X\bm{k}_X}(z_e,z_h),
\end{align}
where
$\phi^X_{\bm{k}}(z_e,z_h)=\chi_e(z_e)\chi_h(z_h)\varphi^{\lambda_X}_{\bm{k}}$
is the in-plane Fourier transform of the exciton envelope function.
We can now construct the state of a single exciton with an in-plane
momentum $\bm{k}_X$ in the Fermionic Hilbert space of electron-hole
pairs. It is the superposition of
wavefunctions~\eqref{eq:X_wavefct_rel_coor} with all electron
momenta $\bm{k}_e$ and all electron $z_e$ and hole $z_h$
coordinates, given by
\begin{align}\label{eq:X_state}
\left|\bm{k}_X\right>=&\sum_{\bm{k}_e} \int \ud z_e \, \ud z_h\,
\phi^{X*}_{\alpha_X\bm{k}_X+\bm{k}_e}(z_e,z_h)\nonumber\\
&\times\hat{c}^\dagger_{-\bm{k}_e,z_e}
\hat{d}^\dagger_{\bm{k}_X+\bm{k}_e,z_h}\left|0\right>,
\end{align}
where $\hat{c}^\dagger_{\bm{k}_X,z_e}$
($\hat{d}^\dagger_{\bm{k}_X,z_h}$) is the electron (hole) creation
operator with in-plane momentum $\bm{k}_X$ and $z_e$ ($z_h$)
coordinate.

Similarly, we choose plane waves for the free carriers. Thus the
in-plane Fourier transform $\psi^\alpha_{\bm{k}}(z)$ of the carrier
wavefunction takes the simple form
\begin{equation}\label{eq:free_career_wave_fct}
\psi^\alpha_{\bm{k}}(z)=\chi_\alpha(z),\quad \alpha=e,h.
\end{equation}
The unbound electron-hole pair then reads
\begin{gather}\label{eq:eh_state}
\left|\bm{k}_e,\bm{k}_h\right>=\int \ud z_e \, \ud \, z_h\,
\psi^{e*}_{\bm{k}_e}(z_e) \,\psi^{h*}_{\bm{k}_h}(z_h)
 \hat{c}^\dagger_{\bm{k}_e,z_e} \hat{d}^\dagger_{\bm{k}_h,z_h}
 \left|0\right>.
\end{gather}

\subsection{Carrier-phonon interaction Hamiltonian.}%
We write the interaction Hamiltonian for a coupled
electron-phonon system in the notation of the second quantization
\citep{mahan:2000}
\begin{equation} \label{eq:Vph}
    \mathcal{H}_{\alpha-ph}
    =  \sum_{\bm{q}, q_z} V^{\alpha}_{\bm{q},q_z}
       \left(
       \hat{a}_{\bm{q}, q_z}
       + \hat{a}_{{-\bm{q}}, {-q}_z}^{\dagger}
       \right)
         \hat{\varrho}_\alpha({\bm{q}},{q}_z)
\end{equation}
where $\hat{a}_{\bm{q},q_z}^{\dagger}$ is the phonon creation
operator. The electron density operator
$\hat{\varrho}_\alpha(\bm{r}_\parallel,z)$ and its counterpart in
Fourier space $\hat{\varrho}_\alpha(\bm{q},q_z)$ are expressed on
the basis $\left\{\phi^\sigma_{\bm{k}}(\bm{r}_{\parallel},z,s) =
e^{i\bm{k}\cdot\bm{r}_{\parallel}}\delta(z)\zeta_\sigma(s)\right\}$
\begin{align}
   \hat{\varrho}_e^\sigma(\bm{r}_\parallel,z)
      & = \sum_{\bm{k},\bm{k}'}
        {\hat{c}_{\bm{k},z}}^{\sigma\dagger}
        \hat{c}_{\bm{k}',z}^{\sigma\phantom{\dagger}}
        e^{-i(\bm{k}-\bm{k}')\cdot\bm{r}_{\parallel}}, \\
   \hat{\varrho}^\sigma_e(\bm{q},q_z)
      & = \sum_{\bm{k}} \int \ud z\, e^{-i q_z z}
        {\hat{c}_{\bm{k}+\bm{q},z}}^{\sigma\dagger}
        \hat{c}_{\bm{k},z}^{\sigma\phantom{\dagger}},
         \\
        \hat{\varrho}_h^\sigma(\bm{r}_\parallel,z)
      & = -\sum_{\bm{k},\bm{k}'}
        {\hat{d}_{\bm{k},z}}^{\sigma\dagger}
        \hat{d}_{\bm{k}',z}^{\sigma\phantom{\dagger}}
        e^{-i(\bm{k}-\bm{k}')\cdot\bm{r}_{\parallel}}, \\
           \hat{\varrho}^\sigma_h(\bm{q},q_z)
      & = -\sum_{\bm{k}} \int \ud z\, e^{-i q_z z}
        {\hat{d}_{\bm{k}+\bm{q},z}}^{\sigma\dagger}
        \hat{d}_{\bm{k},z}^{\sigma\phantom{\dagger}}.
\end{align}
Spin states $\zeta_\sigma(s)=\left<\sigma|s\right>$ have been
intruduced for their will be necessary when we treat the trion
formation.

Only longitudinal acoustical (LA) and longitudinal optical phonons
(LO) couple significantly to careers. We express the coupling vertex
functions $V_{\bm{q},q_z}^{\alpha}$ for both coupling
\begin{align}
V_{\bm{q},q_z}^{\alpha\,(LA)}&=i a_\alpha\sqrt{\frac{\hbar
(|\bm{q}|^2+q_z^2)}{2\rho_0 V \omega_{\bm{q},q_z}}}, \\
V_{\bm{q},q_z}^{\alpha\,(LO)}&=\sqrt{\frac{2\pi\hbar\omega_{\bm{q},q_z}e^2}{(|\bm{q}|^2+q_z^2)V}\left(\frac{1}{\epsilon_0}-\frac{1}{\epsilon_{\infty}}\right)},
\end{align}
where $\epsilon_0$ is the static dielectric constant and
$\epsilon_{\infty}$ is the \emph{high frequency} dielectric
constant. We use the notation $a_\alpha$ for the
deformation-potential constant (assumed to be associated with a
non-degenerate conduction or valence band), $\rho_0$ for the density
of the crystal, $e$ for the charge of the electron and $V$ for the
volume of the sample. We follow Einstein interpolation scheme, so
that the dispersion is merely
$\omega^{LA}_{\bm{q}}=v_s\sqrt{\bm{q}^2+q_z^2}$ for LA phonons and
$\omega_{\bm{q}}^{LO}=\omega_{LO}$ for LO phonons, $v_s$ standing
for the Debye sound velocity and $\omega_{LO}$ for the reststrahl
frequency.

\subsection{Matrix element calculation.}
The matrix elements in Eq. \eqref{eq:fermi_golden_rule} are
calculated, making use of Eq.~\eqref{eq:X_state}, \eqref{eq:eh_state} and~\eqref{eq:Vph}
\begin{widetext}
\begin{multline*}
   \langle\bm{k}_X, n_{\bm{q},q_z} \pm 1|
      \, \mathcal{H}_{\textrm{e-ph}}+\mathcal{H}_{\textrm{h-ph}} \, |\bm{k}_e, \bm{k}_h, n_{\bm{q},q_z}\rangle\\
   \shoveleft{
      = \sum_{\tilde{\bm{q}}, \tilde{q}_z} \sum_{\bm{k},\bm{k}_e'}
        \int \ud{z_e}\ud{z_h}\ud{z_e'}\ud{z_h'}\ud{z}
        \varphi^X_{\alpha \bm{k}_X + \bm{k}_e'}
      \langle n_{\bm{q},q_z}\pm 1|
         \hat{a}_{\pm\tilde{{q}}, \tilde{q}_z}
         \hat{a}^\dagger_{-\tilde{\bm{q}}, -\tilde{q}_z}
         |n_{\bm{q},q_z} \rangle
      \chi_e^*(z_e) \chi_e(z_e')\chi_h^*(z_h) \chi_h(z_h')
      e^{-i\tilde{q}_z z}
   }\\
      \times\Big\{
         V^e_{\tilde{\bm{q}},\tilde{q}_z}
         \langle 0|
            \hat{c}_{-\bm{k}_e',z_e'} \hat{c}_{\bm{k}+\tilde{\bm{q}},z}^\dagger
            \hat{c}_{\bm{k},z} \hat{c}^\dagger_{\bm{k}_e,z_e}
            |0 \rangle
         \langle 0|
            \hat{d}_{\bm{k}_X+\bm{k}_{e}',z_{h}'}
            \hat{d}^\dagger_{\bm{k}_h,z_h}
            |0 \rangle
      - V^h_{\tilde{\bm{q}},\tilde{q}_z}
        \langle 0|
            \hat{c}_{-\bm{k}_e',z_e'}
            \hat{c}^\dagger_{\bm{k}_e,z_e}
            |0 \rangle
         \langle 0|
            \hat{d}_{\bm{k}_X+\bm{k}_{e}',z_{h}'}  \hat{d}_{\bm{k}+\tilde{\bm{q}},z}^\dagger
            \hat{d}_{\bm{k},z} \hat{d}^\dagger_{\bm{k}_h,z_h}
            |0 \rangle
      \Big\}.
\end{multline*}
Applying operators on the ground state
\begin{flalign*}
   &\langle 0|
      \hat{c}_{-\bm{k}_e',z_e'} \hat{c}_{\bm{k}+\tilde{\bm{q}},z}^\dagger
      \hat{c}_{\bm{k},z} \hat{c}^\dagger_{\bm{k}_e,z_e}
      |0 \rangle
   = \delta_{-\bm{k}_e',\bm{k}+\tilde{\bm{q}}} \, \delta(z-z_e')\,
      \delta_{\bm{k},\bm{k}_e} \, \delta(z-z_e), &
   &\langle 0|
      \hat{c}_{-\bm{k}_e',z_e'}
         \hat{c}^\dagger_{\bm{k}_e,z_e}
         |0 \rangle
   = \delta_{\bm{k}_e,-\bm{k}_{e}'} \, \delta(z_e-z_e'),\\
   &\langle 0|
      \hat{d}_{\bm{k}_X+\bm{k}_{e}',z_{h}'}  \hat{d}_{\bm{k}+\tilde{\bm{q}},z}^\dagger
      \hat{d}_{\bm{k},z} \hat{d}^\dagger_{\bm{k}_h,z_h}
      |0 \rangle
   = \delta_{\bm{k}_X+\bm{k}_e',\bm{k}+\tilde{\bm{q}}} \, \delta(z-z_h')\,
      \delta_{\bm{k},\bm{k}_h} \, \delta(z-z_{h}), &
   &\langle 0|
      \hat{d}_{\bm{k}_X+\bm{k}_e',z_h'} \hat{d}^\dagger_{\bm{k}_h,z_h}
      |0 \rangle
   = \delta_{\bm{k}_h,\bm{k}_X+\bm{k}_{e}'} \, \delta(z_h-z_{h}'),\\
   &\langle n_{\bm{q},q_z}\pm 1|
      \hat{a}_{\tilde{\bm{q}},\tilde{q}_z}+ \hat{a}^\dagger_{-\tilde{\bm{q}},-\tilde{q}_z}
      |n_{\bm{q},q_z} \rangle
   = \sqrt{n_{\bm{q},q_z}+\tfrac{1}{2}\pm\tfrac{1}{2}} \; \delta_{\mp\tilde{\bm{q}},\bm{q}} \,
      \delta_{\mp\tilde{q}_z,q_z},\\
\end{flalign*}
we obtain
\begin{multline} \label{eq:matrix_element}
   \langle\bm{k}_X, n_{\bm{q},q_z} \pm 1|
      \, \mathcal{H}_{\textrm{e-ph}}+\mathcal{H}_{\textrm{h-ph}} \, |\bm{k}_e, \bm{k}_h, n_{\bm{q},q_z}\rangle\\
    = \sqrt{n_{\bm{q},q_z}+\tfrac{1}{2}\pm\tfrac{1}{2}} \; \delta_{\pm\bm{q},\bm{k}_e+\bm{k}_h-\bm{k}_X}\Big\{
        V^e_{\bm{q},q_z} \varphi^X_{-\beta_X \bm{k}_X + \bm{k}_h} I_e(q_z)
      - V^h_{\bm{q},q_z} \varphi^X_{\alpha_X \bm{k}_X - \bm{k}_e} I_h(q_z)
      \Big\},
\end{multline}
where the integrals, in the orthogonal direction are given by
\begin{align} \label{eq:Ie}
   I_\alpha(q_z)
   = \int \ud{z_\alpha} |\chi_\alpha(z_\alpha)|^2 e^{iq_z z_\alpha},\qquad\alpha=\{e,\,h\}. 
\end{align}
Finally, if we choose the bound and unbound electron-hole pairs
dispersion to be parabolic, the probability
transition~\eqref{eq:fermi_golden_rule} reads
\begin{multline}\label{fgr2}
   w^\pm_{\bm{k}_e,\bm{k}_h \to \bm{k}_{X}}
   =  \frac{2\pi}{\hbar} \sum_{\bm{q},q_z}
         (n_{\bm{q},q_z}+\tfrac{1}{2}\pm\tfrac{1}{2})\,
      \left|
            V^e_{\bm{q},q_z}\varphi^X_{\beta_X \bm{k}_X - \bm{k}_h} I_e(q_z)
         -  V^h_{\bm{q},q_z}\varphi^X_{\alpha_X \bm{k}_X - \bm{k}_e} I_h(q_z)
       \right|^2\\
   \times
      \delta\left[
      \frac{\hbar^2 k_e^2}{2m_e} + \frac{\hbar^2 k_h^2}{2m_h}
   +  E_b - \frac{\hbar^2k_X^2}{2M} \mp \hbar\omega_{ph}(\bm{q},q_z)
   \right] \delta_{\pm\bm{q},\bm{k}_e+\bm{k}_h-\bm{k}_X}.
\end{multline}
\end{widetext}

\subsection{LA phonons assisted formation}

Considering that the sample volume $V=L_zS$ is macroscopic the sum
over the orthogonal phonon wavevectors may be replaced by the
integral
\begin{align}\label{intqz}
\sum_{q_z} \longleftrightarrow \quad
\left(\frac{L_z}{2\pi}\right)\int \ud q_z 
\end{align}
and reexpressing the Dirac distribution as
\begin{widetext}
\begin{multline}
   \delta\left[
   \frac{\hbar^2 k_e^2}{2m_e} + \frac{\hbar^2 k_h^2}{2m_h}
   +  E_b - \frac{\hbar^2k_X^2}{2M} \mp \hbar v_s \sqrt{\bm{q}^2+q_z^2}
   \right]\\
   = \frac{\sqrt{\bm{q}^2+q_z^2}}{|\hbar v_s q_z|}
     \left\{
       \delta \left[q_z-q_z^{(0)}(\bm{k}_e,\bm{k}_h,\bm{k}_X))\right] +
       \delta \left[q_z+q_z^{(0)}(\bm{k}_e,\bm{k}_h,\bm{k}_X) \right]
     \right\}\;\theta\left[\pm \left(\frac{\hbar^2 k_e^2}{2m_e} + \frac{\hbar^2 k_h^2}{2m_h}
   +  E_b - \frac{\hbar^2k_X^2}{2M} \right)\right],
\end{multline}
with
\begin{equation}
q_z^{(0)}(\bm{k}_e,\bm{k}_h,\bm{k}_X)
 = \sqrt{ \frac{1}{\hbar^2v_s^2}
   \left(
     \frac{\hbar^2 k_e^2}{2m_e} + \frac{\hbar^2 k_h^2}{2m_h}
     +  E_b - \frac{\hbar^2k_X^2}{2M}
   \right)^2 -q^2},
\end{equation}
\end{widetext}
makes the integration \eqref{intqz} trivial for LA phonons:
\begin{multline}\label{FLA}
   w_{\bm{k}_e,\bm{k}_h \to \bm{k}_{X}}
   =
   \frac{4\pi}{\hbar} \frac{L_z}{2\pi}\frac{\hbar}{2\rho V
   v_s} \frac{q^2+q_z^{(0)^2}}{|\hbar v_s q_z^{(0)}|}
       \\
         {\times
      \left|
            a_e\varphi^X_{\beta_X \bm{k}_X - \bm{k}_h} I_e(q_z^{(0)})
         -  a_h\varphi^X_{\alpha_X \bm{k}_X - \bm{k}_e} I_h(q_z^{(0)})
       \right|^2}\\
       \shoveleft{\times \left\{
      \left(n_{\bm{q},q_z^{(0)}}+1\right)\;\theta \left(\frac{\hbar^2 k_e^2}{2m_e} + \frac{\hbar^2 k_h^2}{2m_h}
   +  E_b - \frac{\hbar^2k_X^2}{2M} \right)\right.}\\
   +\left.n_{\bm{q},q_z^{(0)}}\;\theta \left(-\frac{\hbar^2 k_e^2}{2m_e} - \frac{\hbar^2 k_h^2}{2m_h}
   -  E_b + \frac{\hbar^2k_X^2}{2M} \right)\right\}.
\end{multline}
This expression already includes the sum over absorbed and emitted
phonon contributions. The phonon in-plane momentum needs to be
substituted by $\bm{q}=\bm{k}_e+\bm{k}_h-\bm{k}_X$.

\subsection{LO phonons assisted formation}

In the case of interaction with LO phonons, Eq.~\ref{fgr2} becomes
\begin{multline}\label{FLO}
   w_{\bm{k}_e,\bm{k}_h \to \bm{k}_{X}}
   =  \sum_{q_z}\frac{2\pi}{\hbar} \frac{\hbar\omega_{LO}e^2(1/\epsilon_\infty-1/\epsilon_0)}{V(q^2+q_z^2)}
         n_{\bm{q},q_z}\\
         \times 
      \left|
            \varphi^X_{\beta_X \bm{k}_X - \bm{k}_h} I_e(q_z)
         -  \varphi^X_{\alpha_X \bm{k}_X - \bm{k}_e} I_h(q_z)
       \right|^2\\
   \times\delta\left[
      \frac{\hbar^2 k_e^2}{2m_e} + \frac{\hbar^2 k_h^2}{2m_h}
   +  E_b - \frac{\hbar^2k_X^2}{2M} + \hbar\omega_{LO}
   \right],
\end{multline}
where we dropped the phonon absorption part, which is negligible up
to room temperature.
For the calculation of the exciton formation coefficient $C$, it is
convenient to rewrite the Dirac distribution as
\begin{multline}
   \delta\left[
   \frac{\hbar^2 k_e^2}{2m_e} + \frac{\hbar^2 k_h^2}{2m_h}
   +  E_b - \frac{\hbar^2k_X^2}{2M} + \hbar \omega_{LO}
   \right]\\
   = \frac{M_X}{\hbar}
     \left\{
       \delta \left[k_X-k_X^{(0)}(\bm{k}_e,\bm{k}_h))\right] +
       \delta \left[k_X+k_X^{(0)}(\bm{k}_e,\bm{k}_h) \right]
     \right\},
\end{multline}
with
\begin{equation}
k_X^{(0)}(\bm{k}_e,\bm{k}_h)
 = \sqrt{ \frac{2M_X}{\hbar^2}
   \left(
     \frac{\hbar^2 k_e^2}{2m_e} + \frac{\hbar^2 k_h^2}{2m_h}
     +  E_b - \hbar\omega_{LO}.
   \right)},
\end{equation}

\subsection{Numerical Results.}

\begin{figure}
  \includegraphics[width=\linewidth]{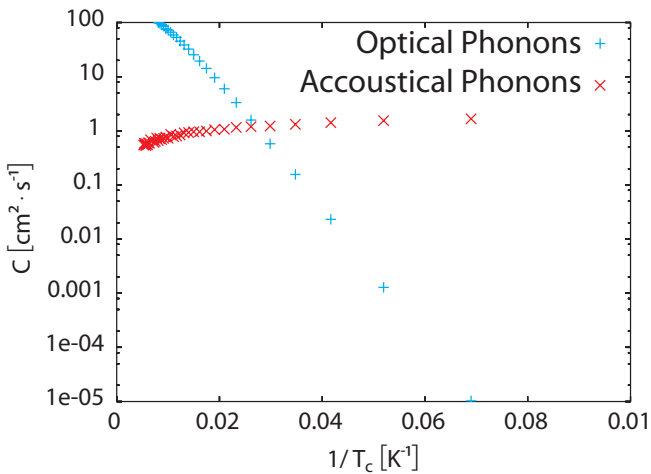}\\
  \caption{The exciton formation coefficient $C$ as a function of the
carrier temperature $T_c$, at a fixed lattice temperature
$T_l=10$~K. Other parameters are given in the
text.}\label{fig:Xformation}
\end{figure}

If we change all the sum in Eq.~\ref{eq:bimol_form_adiab},
\ref{eq:coef_C}, \ref{FLA} and~\ref{FLO} into integrals, the
bimolecular formation coefficient $C$ can be numerically calculated
by Monte Carlo integration. In Fig.~\ref{fig:Xformation}, we report
$C$ as a function of $1/T_c$ for a fixed lattice temperature
$T_l=5$~K, for a GaAs QW of $80$~\AA. The two contributions from the
acoustic and optical phonons are shown separately. The acoustical
phonon dominates for temperatures smaller than $40$~K and does not
depend on $T_l$. We see that these results perfectly match those
published by \citet{piermarocchi:1997}. We show in
Table~\ref{table:GaAs} the numerical value of the different
parameters entering in the calculation.

\begin{table}
  \begin{center}
    \begin{tabular}{l l r l}%
    \hline
    \textbf{Parameter} & \textbf{Symbol} & \textbf{Value} & \textbf{Unit}\\
    \hline
    Band gap energy&$E_g$&$1519$&meV\\%
    Electron effective mass & $m_e$&$0.08$&m$_0$\\%
    Heavy hole effective mass & $m_{hh}$&$0.17$&m$_0$\\%
    LO phonon energy &$\hbar\omega_{LO}$&$36$&meV\\%
    Static dielectric constant & $\epsilon_{0}$&$12.85$&\\%
    High frequency dielectric constant & $\epsilon_{\infty}$&$10.88$&\\%
    Crystal density & $\rho$ & $5.3162$ & g$\cdot$cm$^{-3}$\\%
    Sound velocity & $v_s$&$4726.5$&m$\cdot$s$^{-1}$\\%
    Conduction band deformation potential & $a_e$&$-7.0$&eV\\%
    Valence band deformation potential & $a_h$&$3.5$&eV\\%
    Exciton binding energy & $E_T$&$6.5$&meV\\%
    Trion binding energy & $E_T$&$1.77$&meV\\%
    Exciton Bohr radius & $\lambda_X$ & 11.0 &nm\\ %
    Trion variational parameter \#1 ($X^-$) & $\lambda_T$ & $15$ & nm\\
    Trion variational parameter \#2 ($X^-$) & $\lambda_T'$ & $30.0$ & nm\\
    Trion variational parameter \#1 ($X^+$) & $\lambda_T$ & $16$ & nm\\
    Trion variational parameter \#2 ($X^+$) & $\lambda_T'$ & $25.0$ & nm\\
    \hline
    \end{tabular}
  \end{center}
  \caption{GaAs Material
  Parameters}\label{table:GaAsMaterielParameters}\label{table:GaAs}
\end{table}

\section{Bimolecular formation of trions}
We now extend our formalism to the derivation of the bimolecular
formation of trions. We restrict ourselves to negatively charged
excitons and will give at the end of this work some indication on
how to retrieve their positive counterpart.

We write the scattering term in the Boltzmann equation process for
the bimolecular formation of trions:
\begin{equation} \label{eq:bimol_formT}
\left( \frac{\ud f_e(\bm{k_e})}{\ud{t}}
\right)_{form}=-\sum_{\bm{k_T},\bm{k_e}}
F_{\bm{k_X},\bm{k_e}\to\bm{k_{T}}} f_X(\bm{k_X}) f_e(\bm{k_e}),
\end{equation}
where $F_{\bm{k_h},\bm{k_e}\to\bm{k_{T}}}$ represents the
probability per unit time for a bound electron-hole pair and a free
electron to bind together and form an exciton. We assume that bound
and unbound carriers are thermalized and use Boltzmann distribution
function $f_X(k_X)$ and $f_e(k_e)$ for exciton and electron
population, respectively. By summing Eq.~\eqref{eq:bimol_formT} over
$\bm{k_e}$, we obtain an adiabatic equation for the evolution of the
electron density 
\begin{equation} \label{eq:trimol_form_adiabT}
\frac{\ud n_e}{\ud t}=-\sum_{\bm{k_{T}}} F(\bm{k_{T}}) n_X n_e
\equiv -A_2^- n_X n_e.
\end{equation}
The coefficient $A_2^-$ is the bimolecular formation coefficient,
which depends on both $T_c$ and the lattice temperature $T_l$
through the term
\begin{align} \label{eq:coef_A2Tp}
F(\bm{k_{T}})=&\left( \frac{2 \pi \hbar^2}{k_BT_c} \right)^2
\frac{1}{m_X m_e S} \sum_{\bm{k_X},\bm{k_e}}
w_{\bm{k_X},\bm{k_e}\to\bm{k_{T}}
}\\ &\times e^{-(E_X(\bm{k_X})+E_e(\bm{k_e}))/k_BT_c}.
\end{align}
We calculate the formation rate using Fermi's golden rule
\begin{multline}\label{eq:fermi_golden_ruleT}
w^{\pm}_{\bm{k}_X,\bm{k}_e \to \bm{k}_{T}} = \frac{2\pi}{\hbar}
\sum_{\bm{q},q_z}\\ \left| {\langle\bm{k}_T| \otimes \langle
n_{\bm{q},q_z}\pm1|\, \mathcal{H}_{e/h-ph} \, |n_{\bm{q},q_z}
\rangle \otimes |\bm{k}_X \rangle} \otimes | \bm{k}_e\rangle
 \right|^2\\
 \times \delta\big[ E_X(\bm{k}_X)+E_e(\bm{k}_e)-E_T(\bm{k}_T)\mp\hbar
 \omega_{ph}(\bm{q},q_z)\big],
\end{multline}
with $E_X(\bm{k}_X)$, $E_e(\bm{k}_e)$ and $E_T(\bm{k}_T)$ the energy
dispertion of the electrons, holes and excitons respectively, and
$\hbar \omega_{ph}(\bm{q},q_z)$ the energy of the emitted (absorbed)
phonon. We first build the bound and unbound electron-hole pair
states $|\bm{k}_T \rangle$ and $|\bm{k}_X \rangle \otimes |\bm{k}_e
\rangle$.

\subsection{Trion state.} The two electrons and the hole are
positioned at $(\bm{r}_{1\parallel},z_1)$,
$(\bm{r}_{2\parallel},z_2)$ and $(\bm{r}_{h\parallel},z_h)$
respectively, while the center-of-mass (CM) and relative coordinates
in the QW plane are now given by $\bm{R}_\parallel=\alpha_T
(\bm{r}_{1\parallel}+ \bm{r}_{2\parallel})+
\beta_T\bm{r}_{h\parallel}$ and
$\bm{r}_{i\uh\parallel}=\bm{r}_{i\parallel}-\bm{r}_{h\parallel}$
($i=1,2$); $\alpha_T=m_e/M_T$, $\beta_T=m_h/M_T$ and $M_T$ is the
trion mass. We consider a simple two parameter
Chandrasekhar-type trial envelope function that was successively
used to calculate trion-electron scattering \cite{berney:2008}:
\begin{align}\label{eq:Tenvfct}
\phi&^T(r_{1\uh\parallel},r_{2\uh\parallel},z_1,z_2,z_h) \nonumber\\
=\,&\mathcal{N}_T\chi_e(z_1)\chi_e(z_2)\chi_h(z_h)\nonumber\\
&[\varphi^{\lambda_T}(r_{1\uh})\varphi^{\lambda_T'}(r_{2\uh})\pm
\varphi^{\lambda_T'}(r_{1\uh})\varphi^{\lambda_T}(r_{2\uh})],
\end{align}
where the $+$ ($-$) sign applies to the singlet (triplet) spin
configuration and the trion wavefunction normalization factor is
given by
\begin{equation}
\mathcal{N}_T=\frac{1}{\sqrt{2(1\pm\kappa^2)}},
\end{equation}
with
\begin{equation}
\kappa=\frac{4\lambda\lambda'}{(\lambda+\lambda')^2}.
\end{equation}
Its in-plane Fourier transform reads
\begin{align}\label{eq:T}
&\Phi^T_{\bm{k}_1,\bm{k}_2,\bm{k}_h}(z_1,z_2,z_h)\notag\\
&\quad=\delta_{\bm{k}_T-\bm{k}_1-\bm{k}_2-\bm{k}_h}\,\phi^T_{\alpha_T\bm{k}_T-\bm{k}_1,\alpha_T\bm{k}_T-\bm{k}_2}(z_1,z_2,z_h)
\end{align}
and the state of a single trion with an in-plane CM momentum
$\bm{k}_T$ is constructed similarly to that of a single exciton
\begin{multline}
\left|\bm{k}^S_T\right>=\sum_{\substack{\bm{k}_1,\bm{k}_2\\s_1,s_2}}
\int \ud z_1 \,\ud z_2 \, \ud z_h\,
\phi^{T*}_{\alpha_T\bm{k}_T+\bm{k}_1,\alpha_T\bm{k}_T+\bm{k}_2}(z_1,z_2,z_h)\\
\times \xi^*_S(s_1,s_2) \, \hat{c}^{s_1\dagger}_{-\bm{k}_1,z_1}
\hat{c}^{s_2\dagger}_{-\bm{k}_2,z_2}
\hat{d}^\dagger_{\bm{k}_T+\bm{k}_1+\bm{k}_2,z_h}\left|0\right>,
\end{multline}
where we have added the spin index to the electron creation operator
and introduced $\xi_S(s_1,s_2)=\left<S|s_1,s_2\right>$ the
projection of a generic spin configuration of two electrons on the
singlet spin configuration. The in-plane Fourier transform of the
trion singlet wavefunction in Eq. \eqref{eq:T} is given by
\begin{align}\label{eq:Tft}
\phi^{T*}_{\bm{k}_1,\bm{k}_2}(z_1,z_2,z_h)=&\mathcal{N}_T\chi_e(z_1)\chi_e(z_2)\chi_h(z_h)\nonumber\\
&\times\left[\varphi^{\lambda_T}_{\bm{k}_1}\varphi^{\lambda_T'}_{\bm{k}_2}
+\varphi^{\lambda_T'}_{\bm{k}_1}\varphi^{\lambda_T}_{\bm{k}_2}\right],
\end{align}
where $\varphi^{\lambda_T'}_{\bm{k}_1}$ has already been defined in
Eq.~\ref{eq:1s_fourier}. The exciton-free electron state is given by
 \begin{align}
\left|\bm{k}_X,\bm{k}_{e'}\right>=&\sum_{\bm{k}_e} \int \ud{z_e}
\ud{z_h} \ud{z_{e}'}
\phi^*_{\alpha_X\bm{k}_X+\bm{k}_e}(z_e,z_h) \,\psi^*_{\bm{k}_{e'}}(z_{e'})\notag\\
&\times \, \hat{c}^\dagger_{-\bm{k}_e,z_e}
\hat{d}^\dagger_{\bm{k}_X+\bm{k}_e,z_h}
\hat{c}^\dagger_{\bm{k}_{e'},z_{e'}}\left|0\right>.
\end{align}

\subsection{Matrix element calculation.}
We calculate the matrix elements in \eqref{eq:fermi_golden_ruleT}
\begin{widetext}
\begin{multline}\label{T1}
\langle \bm{k}^S_T;s_h' | \otimes \langle
n_{\bm{q},q_z}\pm1|\,\mathcal{H}_{e/h-ph}\,|n_{\bm{q},q_z}+1\rangle
\otimes |\bm{k}_X;s_1;s_h\rangle \otimes |\bm{k}_2;s_2\rangle
\\
  \shoveleft{=
      \sum_{\bm{k},s} \sum_{\bm{k}_1}
      \sum_{\bm{k}_1',\bm{k}_2'}
      \sum_{s_1',s_2'}
      \int \ud z \,\ud z_1 \, \ud z_2 \, \ud z_h\, \ud z_1' \,\ud z_2' \, \ud z_h'
        \sqrt{n_{\bm{q},q_z}+1}} \\
  {
      \times \xi^*_S(s_1',s_2') \phi^{X}_{\alpha_X\bm{k}_X+\bm{k}_1}(z_1,z_h) \psi_{\bm{k}_2}(z_2)
      \phi^{T*}_{\alpha_T\bm{k}_T+\bm{k}_1',\alpha_T\bm{k}_T+\bm{k}_2'}(z_1',z_2',z_h')} e^{iq_z z} \\
\times \Big[ V_{\bm{q},q_z}^e
    \left<0\right|
        \hat{c}^{\phantom{\dagger}}_1 \hat{c}^{\phantom{\dagger}}_2
        \hat{c}^{\dagger}_3 \hat{c}^{\phantom{\dagger}}_4
        \hat{c}^{\dagger}_5 \hat{c}^{\dagger}_6
    \left|0\right>
    \left<0\right|
        \hat{d}^{\phantom{\dagger}}_7 \hat{d}^{\dagger}_{8}
    \left|0\right>
-   V_{\bm{q},q_z}^h
    \left<0\right|
        \hat{c}^{\phantom{\dagger}}_1 \hat{c}^{\phantom{\dagger}}_2
        \hat{c}^{\dagger}_5 \hat{c}^{\dagger}_6
    \left|0\right>
    \left<0\right|\hat{d}^{\phantom{\dagger}}_7 \hat{d}^{\dagger}_3
        \hat{d}^{\phantom{\dagger}}_4 \hat{d}^{\dagger}_{8}
    \left|0\right>
\Big].
\end{multline}

where we simplified the operators index using the following scheme
\begin{align*}
1 &= (-\bm{k}_1',z_1',s_1') & %
2 &= (-\bm{k}_2',z_2',s_2') &
3 &= (\bm{k}+\bm{q},z,s)\\
4 &= (\bm{k},z,s) & %
5 &= (-\bm{k}_1,z_1,s_1)&
6 &= (\bm{k}_2,z_2,s_2)\\
7 &= (\bm{k}_T+\bm{k}_1'+\bm{k}_2',z_h',s_h')&
8 &= (\bm{k}_X+\bm{k}_1,z_h,s_h)\\
\end{align*}
Using the electron, holes anti-commutation relations, the Fermi
vacuum expectation value of the operators read
\begin{align}
    \left<0\right|
        \hat{c}^{\phantom{\dagger}}_1 \hat{c}^{\phantom{\dagger}}_2
        \hat{c}^{\dagger}_3 \hat{c}^{\phantom{\dagger}}_4
        \hat{c}^{\dagger}_5 \hat{c}^{\dagger}_6
    \left|0\right>
    &= \delta_{13}(\delta_{25}\delta_{46}-\delta_{26}\delta_{45})
    -  \delta_{23}(\delta_{15}\delta_{46}-\delta_{16}\delta_{45})\\
    \left<0\right|
        \hat{d}^{\phantom{\dagger}}_7 \hat{d}^{\dagger}_{8}
    \left|0\right>
    &= \delta_{78}\\
    \left<0\right|
        \hat{c}^{\phantom{\dagger}}_1 \hat{c}^{\phantom{\dagger}}_2
        \hat{c}^{\dagger}_5 \hat{c}^{\dagger}_6
    \left|0\right>
    &= \delta_{25}\delta_{16}-\delta_{15}\delta_{26}\\
    \left<0\right|\hat{d}^{\phantom{\dagger}}_7 \hat{d}^{\dagger}_3
        \hat{d}^{\phantom{\dagger}}_4 \hat{d}^{\dagger}_{8}
    \left|0\right>
    &= \delta_{37}\delta_{48}
\end{align}
Using the later results in Eq.~\eqref{T1} gives
\begin{multline}
\langle \bm{k}^S_T;s_h' | \otimes \langle
n_{\bm{q},q_z}\pm1|\,\mathcal{H}_{e/h-ph}\,|n_{\bm{q},q_z}\rangle
\otimes |\bm{k}_X;s_1;s_h\rangle \otimes |\bm{k}_2;s_2\rangle\\
  \shoveleft{= \left[\xi^*_S(s_2,s_1) - \xi^*_S(s_1,s_2)\right]\delta_{s_h,s_h'}
      \sqrt{n_{\bm{q},q_z}+1/2\pm 1/2}}\; \sum_{\bm{k}_1} \phi^{X}_{\alpha_X\bm{k}_X+\bm{k}_1} \\
      \times  \Big\{ V_{\bm{q},q_z}^e I_e(q_z)
      [\phi^{T*}_{\alpha_T\bm{k}_T-\bm{k}_2,\alpha_T\bm{k}_T+\bm{k}_1+\bm{q}}
      +\phi^{T*}_{\alpha_T\bm{k}_T-\bm{k}_2+\bm{q},\alpha_T\bm{k}_T+\bm{k}_1}]
      - V_{\bm{q},q_z}^h I_h(q_z)
      \phi^{T*}_{\alpha_T\bm{k}_T-\bm{k}_2,\alpha_T\bm{k}_T+\bm{k}_1}\Big\}\delta_{\bm{q},\bm{k}_X+\bm{k}_2-\bm{k}_T}.
\end{multline}
Finally, for parabolic electron, exciton and trion dispersion, the
probability transition \eqref{eq:fermi_golden_ruleT} is
\begin{multline}\label{eq:FermiGoldenRule2}
   w^\pm_{\bm{k}_X,\bm{k}_2 \to \bm{k}_{T}}
   =  \frac{2\pi}{\hbar} \sum_{\bm{q},q_z}
         (n_{\bm{q},q_z}+\tfrac{1}{2}\pm\tfrac{1}{2})\,\\
      \times\bigg|
            \sum_{\bm{k}_1} V_{\bm{q},q_z}^e I_e(q_z)
                [\phi^{T*}_{\alpha_T\bm{k}_T-\bm{k}_2,\alpha_T\bm{k}_T+\bm{k}_1+\bm{q}}
                +\phi^{T*}_{\alpha_T\bm{k}_T-\bm{k}_2+\bm{q},\alpha_T\bm{k}_T+\bm{k}_1}]
            - V_{\bm{q},q_z}^h I_h(q_z)
                \phi^{T*}_{\alpha_T\bm{k}_T-\bm{k}_2,\alpha_T\bm{k}_T+\bm{k}_1}
       \bigg|^2\\
   \times
      \delta\left[
      \frac{\hbar^2 k_X^2}{2M_X} + \frac{\hbar^2 k_2^2}{2m_e}
   +  E_T - \frac{\hbar^2k_T^2}{2M_T} \mp \hbar\omega_{ph}(\bm{q},q_z)
   \right] \delta_{\pm\bm{q},\bm{k}_X+\bm{k}_2-\bm{k}_T},
\end{multline}
where we averaged over the initial electron spin states and exciton
angular momentum states.

\end{widetext}

\section{Trimolecular formation of trions}

Again, we write the scattering term in the Boltzmann equation
process

\begin{equation} \label{eq:trimol_formT}
\left( \frac{\ud f_e(\bm{k}_e)}{\ud t}
\right)_{form}=-\sum_{\bm{k}_T,\bm{k}_e'}
F_{\bm{k}_e,\bm{k}_e',\bm{k}_h\to\bm{k_{T}}} f_e(\bm{k}_e)
f_e(\bm{k}_e') f_h(\bm{k}_h),
\end{equation}
where $F_{\bm{k}_e,\bm{k}_e',\bm{k}_h\to\bm{k_{T}}}$ represents the
probability per unit time for two free electrons and one free hole
to bind together and form an exciton. We assume that bound and
unbound carriers are thermalized and use Boltzmann distribution
function $f_X(k_e)$ and $f_h(k_h)$ for exciton and electron
population, respectively. By summing Eq.~\eqref{eq:trimol_formT}
over $\bm{k}_e$, we obtain an adiabatic equation for the evolution
of the electron density 
\begin{equation} \label{eq:bimol_form_adiabT}
\frac{\ud{n_e}}{\ud{t}}=-\sum_{\bm{k_{T}}} F(\bm{k_{T}}) n_e^2 n_h
\equiv -A_3^- n_e^2 n_h.
\end{equation}
The coefficient $A_3^-$ is the trimolecular formation coefficient,
which depends on both $T_c$ and the lattice temperature $T_l$
through the term
\begin{align} \label{eq:coef_A3Tp}
F(\bm{k_{T}})=&\left( \frac{2 \pi \hbar^2}{k_BT_c} \right)^3
\frac{1}{m_e^2 m_h S^{3/2}}
\sum_{\bm{k}_e,\bm{k}_e',\bm{k}_h}
w_{\bm{k}_e,\bm{k}_e,\bm{k}_h'\to\bm{k_{T}}
}  \nonumber\\ & \times e^{-(E_e(\bm{k_e})+E_e(\bm{k_e'})+E_h(\bm{k_h}))/k_BT_c}.
\end{align}
We calculate the formation rate using Fermi's golden rule
\begin{widetext}
\begin{multline}\label{eq:fermi_golden_ruleT3}
w^{\pm}_{\bm{k}_e,\bm{k}_e',\bm{k}_h \to \bm{k}_{T}} =
\frac{2\pi}{\hbar} \sum_{\bm{q},q_z} \left| {\langle\bm{k}_T|
\otimes \langle n_{\bm{q},q_z}\pm1|\, \mathcal{H}_{e/h-ph} \,
|n_{\bm{q},q_z} \rangle \otimes |\bm{k}_e \rangle} \otimes |
\bm{k}_e'\rangle \otimes | \bm{k}_h\rangle
 \right|^2\\
 \times \delta\big[ E_e(\bm{k}_e)+E_e(\bm{k}_e')+E_h(\bm{k}_h)-E_T(\bm{k}_T)\mp\hbar
 \omega_{ph}(\bm{q},q_z)\big],
\end{multline}
with $E_X(\bm{k}_X)$, $E_e(\bm{k}_e)$ and $E_T(\bm{k}_T)$ the energy
dispertion of the electrons, holes and excitons respectively, and
$\hbar \omega_{ph}(\bm{q},q_z)$ the energy of the emitted (absorbed)
phonon.
Finally, for parabolic electron, exciton and trion dispersion, the
probability transition \ref{eq:fermi_golden_ruleT3} becomes
\begin{multline}\label{eq:FermiGoldenRule3}
   w^\pm_{\bm{k}_e,\bm{k}_e',\bm{k}_h \to \bm{k}_{T}}
   =  \frac{2\pi}{\hbar} \sum_{\bm{q},q_z}
         (n_{\bm{q},q_z}+\tfrac{1}{2}\pm\tfrac{1}{2})\,\\
      \times\bigg|
            V_{\bm{q},q_z}^e I_e(q_z)
                [\phi^{T*}_{\alpha_T\bm{k}_T-\bm{k}_2,\alpha_T\bm{k}_T+\bm{k}_1+\bm{q}}
                +\phi^{T*}_{\alpha_T\bm{k}_T-\bm{k}_2+\bm{q},\alpha_T\bm{k}_T+\bm{k}_1}]
            - V_{\bm{q},q_z}^h I_h(q_z)
                \phi^{T*}_{\alpha_T\bm{k}_T-\bm{k}_2,\alpha_T\bm{k}_T+\bm{k}_1}
       \bigg|^2\\
   \times
      \delta\left[
      \frac{\hbar^2 k_e^2}{2m_e} + \frac{\hbar^2 k_e'^2}{2m_e} + \frac{\hbar^2 k_h^2}{2m_h}
   +  E_T - \frac{\hbar^2k_T^2}{2M_T} \mp \hbar\omega_{ph}(\bm{q},q_z)
   \right] \delta_{\pm\bm{q},\bm{k}_X+\bm{k}_2-\bm{k}_T},
\end{multline}
where we averaged over the initial electron spin states and exciton
angular momentum states.
\end{widetext}
\subsection{Numerical results}

\begin{figure}[thb]
  \includegraphics[width=\linewidth]{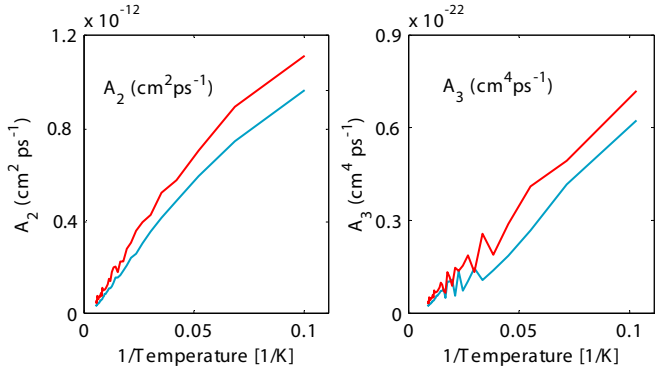}
\caption{\label{fig4} The bi- and tri- molecular trion formation
coefficients $A_2^{\alpha}$ and $A_3^{\alpha}$ as a function of the
inverse carrier temperature as calculated from our model for LA
phonon assisted formation. The red curve indicates $X^+$ formation
and the blue curve $X^-$ formation.}\label{fig:bitri}
\end{figure}

In Fig.~\ref{fig:bitri}, we represent the results of our numerical
calculation for LA phonon assisted formation. We stress the fact
that the results for LO phonons are orders of magnitude smaller and
consequently negligible for bi- and tri-molecular processes. This
shows that the exclusive formation mechanism for trions is governed
by LA phonon interaction. We predict a decrease of the bi-
or tri-molecular formation for raising temperatures. We also
demonstrate that the bi and tri-molecular formation coefficients for
negatively and positively charged excitons are approximatively equal
($A_2^+=A_2^-$) and ($A_3^+=A_3^-$).

A more accurate calculation
should rely on a better trion wavefunction. The Chandrashekar
variational function is most likely to simple to yield quantitative
results. We note however that electron-trion scattering \cite{ramon:2003}
will contribute to ionize trions.
This effect will dramatically increase at high temperatures. We
consequently propose that the electron-trion scattering reduce the
experimental values for the trion formation coefficients.

\section{Formation rates close to equilibrium}
In this Section, we propose to derive formation rate for electrons, holes, excitons and trions assuming thermodynamical equilibrium. In this set of equations, we neglect both biexciton channels and Auger channels because experiments on undoped samples demonstrate that, at the densities considered in the present work, these channels are not significant \cite{szczytko:2004}. 

The dynamics of a plasma of electrons containing electrons ($e$), holes ($h$), excitons ($X$) and trions ($X^+$ and $X^-$)  is governed by the following five channels:

\begin{align*}
e+h & \leftrightarrow X\\
X+e & \leftrightarrow X^- &
X+h & \leftrightarrow X^+\\
2e+h & \leftrightarrow X^- &
2h+e & \leftrightarrow X^+
\end{align*}

The different formation rates for these populations read
\begin{align}
F^{X}&= C \mathbf{n}\mathbf{p}- \gamma C K_X \mathbf{X},\\
F_2^{X^{-}}& =  A_2^- \mathbf{X}\mathbf{n} - A_2^- K_2^{-} \mathbf{X^{-}},\\
F_3^{X^{-}} &=  A_3^- \mathbf{n}\mathbf{n}\mathbf{p} - A_3^- K_3^{-} \mathbf{X^{-}},\\
F_2^{X^{+}} &=  A_2^+ \mathbf{X}\mathbf{p} - A_2^+ K_2^{+} \mathbf{X^{+}},\\
F_3^{X^{+}} &= A_3^+ \mathbf{n}\mathbf{p}\mathbf{p} - A_3^+ K_3^{+} \mathbf{X^{+}},
\end{align}
where $C$, $A_2^\alpha$ and $A_3^\alpha$ are respectively the exciton, trion bimolecular and trion trimolecular formation rate calculated in this article; $E_{bX}$ is the exciton binding energy;$K_X$, $K_2^{\alpha}$, $K_3^{\alpha}$  the equilibrium coefficients. For a 2D system, they can be derived from the Boltzmann distribution
\begin{equation}
n_\alpha=2 g_\alpha \int \frac{d^2k}{2\pi}e^{-hbar^2k^2/2m_\alpha k_BT} =  g_\alpha \frac{m_\alpha k_B T}{2\pi \hbar^2}e^{-\mu_\alpha/K_BT},
\end{equation}
where $\alpha=e, h, X,T$, the factor $g_\alpha=2$ is the spin degeneracy of the electron, hole, exction and trion in the non degenerate regime, where Boltzmann statistics applies.

Using that fact that the chemical potential of the exciton $\mu_X$ is related to the chemical potential of electrons $\mu_e$ and holes $\mu_h$ as $\mu_X=\mu_e+\mu_h+E_X$ ($E_X$ is here defined as minus the exciton binding energy), one immediately obtains the Saha relation for the exciton density
\begin{equation}
\frac{np}{X} = K_X(T)=\frac{g_e g_h}{g_X}\frac{m_e m_h}{m_X}  \frac{ k_B T}{2\pi\hbar^2}e^{{-E_X}/{k_BT}}.
\end{equation}

Similarly, the chemical potential of the trions is $\mu_T=2\mu_e+\mu_h+E_T$ ($E_T$ is here defined as minus the trion binding energy), so that Saha equations for trion bi-molecular formation are
\begin{align}
\frac{Xn}{X^-} &= K_2^-(T)=\frac{g_e^2 g_X}{g_T}\frac{m_X m_e}{m_{X^-}}  \frac{ k_B T}{2\pi\hbar^2}e^{{-(E_T-E_X)}/{k_BT}}\\
\frac{Xp}{X^+}& = K_2^+(T)=\frac{g_h^2 g_X}{g_T}\frac{m_X m_h}{m_{X^+}}  \frac{ k_B T}{2\pi\hbar^2}e^{{-(E_T-E_X)}/{k_BT}}
\end{align}

Finally, the following set of equation is infered for tri-molecular formation:
\begin{align}
\frac{n^2h}{X^-}& = K_3^-(T)=\frac{g_e^2 g_h}{g_T}\frac{m_e^2 m_h}{m_{X^-}} \left( \frac{ k_B T}{2\pi\hbar^2}\right)^{2}e^{{-E_T}/{k_BT}}\\
\frac{nh^2}{X^+}& = K_3^+(T)=\frac{g_e g_h^2}{g_T}\frac{m_e m_h^2}{m_{X^+}}  \left(\frac{ k_B T}{2\pi\hbar^2}\right)^{2}e^{{-E_T}/{k_BT}}
\end{align}

Such a set of equations, together with the equilibrium densities, allow to compute the dynamics of the different populations after non-resonant optical excitation. This may apply both to undoped \cite{szczytko:2004} as well as to doped quantum wells. The results in the case of a sample doped with electrons will be detailed in another publication \cite{portella:2009}. The experiments show that indeed, when the density of electrons is sufficient and at excitation densities of the order of $10^{10}$cm$^{-2}$ and above, the trimolecular formation process of trions has to be taken into account to properly reproduce the observed dynamics.  
 
\section{conclusion}

In this paper, we have derived the appropriate model for computing the rates for exciton and trion formation. For the case of trions, we have derived the equations for both the bi- and tri-molecular phonon-assisted formation of trions. We have shown that bi- and tri-molecular formation rate of negatively and positively charged excitons have similar orders of magnitude for densities that are used in the experiments. We have then developed the set of relations and the equilibrium conditions allowing to calculate the dynamics of free carriers, excitons and trions. Our results are in very reasonable agreement with recent experiments and allow to confirm that, indeed, trimolecular trion formation may not be neglected in real samples.

\section{Acknowledgments}

We wish to thank Fabienne Michelini, Michiel Wouters, Christiano Ciuti, Carlo Piermarocchi and Vicenzo Savona for fruitful discussions. We also thank the Swiss National Fundation for funding.

\bibliography{PRB_formation_v1.0}{}

\begin{thebibliography}{18}
\expandafter\ifx\csname natexlab\endcsname\relax\def\natexlab#1{#1}\fi
\expandafter\ifx\csname bibnamefont\endcsname\relax
  \def\bibnamefont#1{#1}\fi
\expandafter\ifx\csname bibfnamefont\endcsname\relax
  \def\bibfnamefont#1{#1}\fi
\expandafter\ifx\csname citenamefont\endcsname\relax
  \def\citenamefont#1{#1}\fi
\expandafter\ifx\csname url\endcsname\relax
  \def\url#1{\texttt{#1}}\fi
\expandafter\ifx\csname urlprefix\endcsname\relax\def\urlprefix{URL }\fi
\providecommand{\bibinfo}[2]{#2}
\providecommand{\eprint}[2][]{\url{#2}}

\bibitem[{\citenamefont{Weisbuch et~al.}(1981)\citenamefont{Weisbuch, Miller,
  Dingle, Gossard, and Wiegman}}]{weisbuch:1981}
\bibinfo{author}{\bibfnamefont{C.}~\bibnamefont{Weisbuch}},
  \bibinfo{author}{\bibfnamefont{R.~C.} \bibnamefont{Miller}},
  \bibinfo{author}{\bibfnamefont{R.}~\bibnamefont{Dingle}},
  \bibinfo{author}{\bibfnamefont{A.~C.} \bibnamefont{Gossard}},
  \bibnamefont{and} \bibinfo{author}{\bibfnamefont{W.}~\bibnamefont{Wiegman}},
  \bibinfo{journal}{Solid State Communications} \textbf{\bibinfo{volume}{37}},
  \bibinfo{pages}{219 } (\bibinfo{year}{1981}), ISSN \bibinfo{issn}{0038-1098}.

\bibitem[{\citenamefont{Deveaud et~al.}(1987)\citenamefont{Deveaud, Damen,
  Shah, and Tu}}]{deveaud:1987}
\bibinfo{author}{\bibfnamefont{B.}~\bibnamefont{Deveaud}},
  \bibinfo{author}{\bibfnamefont{T.~C.} \bibnamefont{Damen}},
  \bibinfo{author}{\bibfnamefont{J.}~\bibnamefont{Shah}}, \bibnamefont{and}
  \bibinfo{author}{\bibfnamefont{C.~W.} \bibnamefont{Tu}},
  \bibinfo{journal}{Applied Physics Letters} \textbf{\bibinfo{volume}{51}},
  \bibinfo{pages}{828} (\bibinfo{year}{1987}).

\bibitem[{\citenamefont{Yoon et~al.}(1996)\citenamefont{Yoon, Wake, and
  Wolfe}}]{yoon:1996}
\bibinfo{author}{\bibfnamefont{H.~W.} \bibnamefont{Yoon}},
  \bibinfo{author}{\bibfnamefont{D.~R.} \bibnamefont{Wake}}, \bibnamefont{and}
  \bibinfo{author}{\bibfnamefont{J.~P.} \bibnamefont{Wolfe}},
  \bibinfo{journal}{Phys. Rev. B} \textbf{\bibinfo{volume}{54}},
  \bibinfo{pages}{2763} (\bibinfo{year}{1996}).

\bibitem[{\citenamefont{Damen et~al.}(1990)\citenamefont{Damen, Shah, Oberli,
  Chemla, Cunningham, and Kuo}}]{damen:1990}
\bibinfo{author}{\bibfnamefont{T.~C.} \bibnamefont{Damen}},
  \bibinfo{author}{\bibfnamefont{J.}~\bibnamefont{Shah}},
  \bibinfo{author}{\bibfnamefont{D.~Y.} \bibnamefont{Oberli}},
  \bibinfo{author}{\bibfnamefont{D.~S.} \bibnamefont{Chemla}},
  \bibinfo{author}{\bibfnamefont{J.~E.} \bibnamefont{Cunningham}},
  \bibnamefont{and} \bibinfo{author}{\bibfnamefont{J.~M.} \bibnamefont{Kuo}},
  \bibinfo{journal}{Phys. Rev. B} \textbf{\bibinfo{volume}{42}},
  \bibinfo{pages}{7434} (\bibinfo{year}{1990}).

\bibitem[{\citenamefont{Robart et~al.}(1995)\citenamefont{Robart, Marie,
  Baylac, Amand, Brousseau, Bacquet, Debart, Planel, and Gerard}}]{robart:1995}
\bibinfo{author}{\bibfnamefont{D.}~\bibnamefont{Robart}},
  \bibinfo{author}{\bibfnamefont{X.}~\bibnamefont{Marie}},
  \bibinfo{author}{\bibfnamefont{B.}~\bibnamefont{Baylac}},
  \bibinfo{author}{\bibfnamefont{T.}~\bibnamefont{Amand}},
  \bibinfo{author}{\bibfnamefont{M.}~\bibnamefont{Brousseau}},
  \bibinfo{author}{\bibfnamefont{G.}~\bibnamefont{Bacquet}},
  \bibinfo{author}{\bibfnamefont{G.}~\bibnamefont{Debart}},
  \bibinfo{author}{\bibfnamefont{R.}~\bibnamefont{Planel}}, \bibnamefont{and}
  \bibinfo{author}{\bibfnamefont{J.~M.} \bibnamefont{Gerard}},
  \bibinfo{journal}{Solid State Communications} \textbf{\bibinfo{volume}{95}},
  \bibinfo{pages}{287 } (\bibinfo{year}{1995}), ISSN \bibinfo{issn}{0038-1098}.

\bibitem[{\citenamefont{Deveaud}(1993)}]{deveaud:1993}
\bibinfo{author}{\bibfnamefont{B.}~\bibnamefont{Deveaud}}, \bibinfo{journal}{J.
  Phys. IV} \textbf{\bibinfo{volume}{3}}, \bibinfo{pages}{11}
  (\bibinfo{year}{1993}).

\bibitem[{\citenamefont{Kaindl et~al.}(2003)\citenamefont{Kaindl, Carnahan,
  Hagele, Lovenich, and Chemla}}]{kaindl:2003}
\bibinfo{author}{\bibfnamefont{R.~A.} \bibnamefont{Kaindl}},
  \bibinfo{author}{\bibfnamefont{M.~A.} \bibnamefont{Carnahan}},
  \bibinfo{author}{\bibfnamefont{D.}~\bibnamefont{Hagele}},
  \bibinfo{author}{\bibfnamefont{R.}~\bibnamefont{Lovenich}}, \bibnamefont{and}
  \bibinfo{author}{\bibfnamefont{D.~S.} \bibnamefont{Chemla}},
  \bibinfo{journal}{Nature} \textbf{\bibinfo{volume}{423}},
  \bibinfo{pages}{734} (\bibinfo{year}{2003}).

\bibitem[{\citenamefont{Szczytko et~al.}(2004)\citenamefont{Szczytko, Kappei,
  Berney, Morier-Genoud, Portella-Oberli, and Deveaud}}]{szczytko:2004}
\bibinfo{author}{\bibfnamefont{J.}~\bibnamefont{Szczytko}},
  \bibinfo{author}{\bibfnamefont{L.}~\bibnamefont{Kappei}},
  \bibinfo{author}{\bibfnamefont{J.}~\bibnamefont{Berney}},
  \bibinfo{author}{\bibfnamefont{F.}~\bibnamefont{Morier-Genoud}},
  \bibinfo{author}{\bibfnamefont{M.~T.} \bibnamefont{Portella-Oberli}},
  \bibnamefont{and} \bibinfo{author}{\bibfnamefont{B.}~\bibnamefont{Deveaud}},
  \bibinfo{journal}{Phys. Rev. Lett.} \textbf{\bibinfo{volume}{93}},
  \bibinfo{pages}{137401} (\bibinfo{year}{2004}).

\bibitem[{\citenamefont{Szczytko et~al.}(2005)\citenamefont{Szczytko, Kappei,
  Berney, Morier-Genoud, Portella-Oberli, and Deveaud}}]{szczytko:2005}
\bibinfo{author}{\bibfnamefont{J.}~\bibnamefont{Szczytko}},
  \bibinfo{author}{\bibfnamefont{L.}~\bibnamefont{Kappei}},
  \bibinfo{author}{\bibfnamefont{J.}~\bibnamefont{Berney}},
  \bibinfo{author}{\bibfnamefont{F.}~\bibnamefont{Morier-Genoud}},
  \bibinfo{author}{\bibfnamefont{M.~T.} \bibnamefont{Portella-Oberli}},
  \bibnamefont{and} \bibinfo{author}{\bibfnamefont{B.}~\bibnamefont{Deveaud}},
  \bibinfo{journal}{Physical Review B} \textbf{\bibinfo{volume}{71}},
  \bibinfo{pages}{195313} (\bibinfo{year}{2005}).

\bibitem[{\citenamefont{Thilagam and Singh}(1993)}]{thilagam:1993}
\bibinfo{author}{\bibfnamefont{A.}~\bibnamefont{Thilagam}} \bibnamefont{and}
  \bibinfo{author}{\bibfnamefont{J.}~\bibnamefont{Singh}},
  \bibinfo{journal}{Physical Review B} \textbf{\bibinfo{volume}{48}},
  \bibinfo{pages}{4636} (\bibinfo{year}{1993}).

\bibitem[{\citenamefont{Piermarocchi et~al.}(1996)\citenamefont{Piermarocchi,
  Tassone, Savona, Quattropani, and Schwendimann}}]{piermarocchi:1996}
\bibinfo{author}{\bibfnamefont{C.}~\bibnamefont{Piermarocchi}},
  \bibinfo{author}{\bibfnamefont{F.}~\bibnamefont{Tassone}},
  \bibinfo{author}{\bibfnamefont{V.}~\bibnamefont{Savona}},
  \bibinfo{author}{\bibfnamefont{A.}~\bibnamefont{Quattropani}},
  \bibnamefont{and}
  \bibinfo{author}{\bibfnamefont{P.}~\bibnamefont{Schwendimann}},
  \bibinfo{journal}{Phys. Rev. B} \textbf{\bibinfo{volume}{53}},
  \bibinfo{pages}{15834} (\bibinfo{year}{1996}).

\bibitem[{\citenamefont{Piermarocchi et~al.}(1997)\citenamefont{Piermarocchi,
  Tassone, Savona, Quattropani, and Schwendimann}}]{piermarocchi:1997}
\bibinfo{author}{\bibfnamefont{C.}~\bibnamefont{Piermarocchi}},
  \bibinfo{author}{\bibfnamefont{F.}~\bibnamefont{Tassone}},
  \bibinfo{author}{\bibfnamefont{V.}~\bibnamefont{Savona}},
  \bibinfo{author}{\bibfnamefont{A.}~\bibnamefont{Quattropani}},
  \bibnamefont{and}
  \bibinfo{author}{\bibfnamefont{P.}~\bibnamefont{Schwendimann}},
  \bibinfo{journal}{Phys. Rev. B} \textbf{\bibinfo{volume}{55}},
  \bibinfo{pages}{1333} (\bibinfo{year}{1997}).

\bibitem[{\citenamefont{Portella-Oberli
  et~al.}(2009)\citenamefont{Portella-Oberli, Berney, Kappei, Morier-Genoud,
  Szczytko, and Deveaud}}]{portella:2009}
\bibinfo{author}{\bibfnamefont{M.~T.} \bibnamefont{Portella-Oberli}},
  \bibinfo{author}{\bibfnamefont{J.}~\bibnamefont{Berney}},
  \bibinfo{author}{\bibfnamefont{L.}~\bibnamefont{Kappei}},
  \bibinfo{author}{\bibfnamefont{F.}~\bibnamefont{Morier-Genoud}},
  \bibinfo{author}{\bibfnamefont{J.}~\bibnamefont{Szczytko}}, \bibnamefont{and}
  \bibinfo{author}{\bibfnamefont{B.}~\bibnamefont{Deveaud}},
  \bibinfo{journal}{submitted to Phys. Rev. Lett.}  (\bibinfo{year}{2009}).

\bibitem[{\citenamefont{Knox}(1992)}]{knox:1992}
\bibinfo{author}{\bibfnamefont{W.~H.} \bibnamefont{Knox}},
  \emph{\bibinfo{title}{Optical Studies of Femtosecond Carrier Thermalization
  in GaAs}} (\bibinfo{publisher}{Academic Press}, \bibinfo{address}{San Diego},
  \bibinfo{year}{1992}), p. \bibinfo{pages}{313}.

\bibitem[{\citenamefont{Bastard}(1988)}]{bastard:1988}
\bibinfo{author}{\bibfnamefont{G.}~\bibnamefont{Bastard}},
  \emph{\bibinfo{title}{Wave mechanics applied to semiconductor
  heterostrutures}} (\bibinfo{publisher}{Les \'editions de physique},
  \bibinfo{address}{Les Ulis, France}, \bibinfo{year}{1988}).

\bibitem[{\citenamefont{Mahan}(2000)}]{mahan:2000}
\bibinfo{author}{\bibfnamefont{G.~D.} \bibnamefont{Mahan}},
  \emph{\bibinfo{title}{Many Particle Physics}} (\bibinfo{publisher}{Springer},
  \bibinfo{year}{2000}), \bibinfo{edition}{3rd} ed.

\bibitem[{\citenamefont{Berney et~al.}(2008)\citenamefont{Berney,
  Portella-Oberli, and Deveaud}}]{berney:2008}
\bibinfo{author}{\bibfnamefont{J.}~\bibnamefont{Berney}},
  \bibinfo{author}{\bibfnamefont{M.~T.} \bibnamefont{Portella-Oberli}},
  \bibnamefont{and} \bibinfo{author}{\bibfnamefont{B.}~\bibnamefont{Deveaud}},
  \bibinfo{journal}{Physical Review B (Condensed Matter and Materials Physics)}
  \textbf{\bibinfo{volume}{77}}, \bibinfo{eid}{121301}
  (pages~\bibinfo{numpages}{4}) (\bibinfo{year}{2008}).

\bibitem[{\citenamefont{Ramon et~al.}(2003)\citenamefont{Ramon, Mann, and
  Cohen}}]{ramon:2003}
\bibinfo{author}{\bibfnamefont{G.}~\bibnamefont{Ramon}},
  \bibinfo{author}{\bibfnamefont{A.}~\bibnamefont{Mann}}, \bibnamefont{and}
  \bibinfo{author}{\bibfnamefont{E.}~\bibnamefont{Cohen}},
  \bibinfo{journal}{Physical Review B} \textbf{\bibinfo{volume}{67}},
  \bibinfo{pages}{45323} (\bibinfo{year}{2003}).

\end{thebibliography}
\end{document}